\documentclass[aps,prc,twocolumn,showkeys,showpacs,groupedaddress,superscriptaddress]{revtex4}
\usepackage{amssymb}
\usepackage{graphicx}
\usepackage{color}

\newcommand{\req}[1]{(\ref{#1})}
\newcommand{\bel}[1]{\begin{equation}\label{#1}}
\newcommand{\belar}[1]{\begin{eqnarray}\label{#1}}

\begin{document}

\title{Fission of super-heavy elements: $^{132}$Sn-plus-the-rest, or 
$^{208}$Pb-plus-the-rest ?}

\author{C. Ishizuka}
\email{chikako@nr.titech.ac.jp}
\author{X. Zhang}
\email{zhang.x.ba@m.titech.ac.jp}
\affiliation{Tokyo Institute of Technology, Tokyo, 152-8550 Japan}
\author{M.D. Usang}
\email{mark_dennis@nuclearmalaysia.gov.my}
\affiliation{Malaysia Nuclear Agency, Bangi, Malaysia}
\author{F.A. Ivanyuk}
\email{ivanyuk@kinr.kiev.ua}
\affiliation{Tokyo Institute of Technology, Tokyo, 152-8550 Japan}
\affiliation{Institute for Nuclear Research, 03028  Kiev, Ukraine}
\author{S. Chiba}
\email{chiba.satoshi@nr.titech.ac.jp}
\affiliation{Tokyo Institute of Technology, Tokyo, 152-8550 Japan}
\affiliation{National Astronomical Observatory  of Japan, Tokyo, Japan}
\date{today}

\begin{abstract}
In this work we try to settle down the controversial predictions on 
the effect of doubly magic nuclei $^{132}$Sn and 
$^{208}$Pb on the mass distributions of fission fragments of 
super-heavy nuclei. For this we have calculated the mass distribution of fission fragments of
super-heavy nuclei from $^{274}$Hs to $^{306}$122 within 
the dynamical 4-dimensional Langevin approach. We have found that in 
``light'' super-heavies the influence of $^{208}$Pb on the mass 
distributions is negligible small. In ''heavy'' super-heavies, 
$Z=120-122$, the (quasi)symmetric peaks and strongly asymmetric peaks at 
fragment mass $A_F$ close to $A_F=208$ are of 
comparable magnitude to $A_F=132-140$.
\end{abstract}


\maketitle

\section{Introduction}
\label{intro}
The physics of super-heavy elements (SHE) has a long history. The existence 
of the ``island of stability'' -- the region of nuclei with the increased 
stability with respect to spontaneous fission - was predicted at the middle of 1960s. The possibility of closed shells at $Z=114, N=184$ was pointed out already in \cite{swiat,sobicz,meldner}. The systematic calculations in \cite{1} within the  macroscopic-microscopic method \cite{2,3,4} 
for SHEs with the number of protons $106<Z<116$ have shown that many 
super-heavies are very stable, with the spontaneous fission half-lives of 
thousands years or more. The highest fission barrier was predicted for a new 
double magic nucleus with $Z=114$ and $N=184$. Nevertheless, it took almost 30 
years until the alpha-decay of the element with $Z=114$ was observed 
experimentally at Flerov Laboratory for Nuclear Reactions in Dubna \cite{5}. During 
the next two decades a lot of new experimental achievements were synthesized. The theoretical works were 
dedicated to the search of most favorable pairs of projectile and target and 
the excitation energy that would lead to the largest cross section of 
formation of evaporation residue -- the super-heavy nucleus in its ground 
state.

With the development of experimental facility it became possible not only 
fix the fact of formation of SHE, but accumulate so many super-heavy nuclei 
that it turned out possible to examine their properties. One of the first 
property of interest -- the process of fission of SHEs. For the successful 
planning and carrying out experiments it is very important to understand 
what kind of fission fragments one should expect in the result of fission of 
SHE. On one side, it is clear that with increasing charge number $Z$ of 
fissioning nucleus the Coulomb repulsion force grows and one could expect 
the symmetric mass distribution of fission fragments. One other side -- the 
shell effects may still have a noticeable effect. The two double magic 
nuclei may contribute. The $^{132}$Sn and $^{208}$Pb have the shell 
correction to the ground state energy of the same magnitude. The $^{132}$Sn plays a 
decisive role in formation of mass distribution of actinide and 
trans-actinide nuclei. In the experiment of Itkis group \cite{6,7} $^{132}$Sn 
was found out as the light fragment of all investigated nuclei. The 
theoretical calculation within the scission point model \cite{8} also predict 
$^{132}$Sn (or slightly heavier) as the most probably light fragment for 
fission of SHE. At the same time there are few publications \cite{9,10,11,12} where 
formation of heavy fragment close to $^{208}$Pb is predicted as a main 
fission mode. In \cite{13} the heavy fragment close to $^{208}$Pb was obtained 
in the super-heavy region, $106<Z<114$. 

In order to solve this contradiction and make it clear what kind of fission 
fragment mass distribution (FFMD) one could expect in the fission of SHEs, 
we have carried out the calculations of FFMD for a number of SHEs within the 
four-dimensional Langevin approach. We have found out the $^{208}$Pb may 
appear as a supplementary heavy cluster in fission of Cn isotopes. With 
increasing charge number of SHEs the contribution of this heavy cluster 
became larger. For the element with $Z=122$ the contributions of (almost) 
symmetric and strongly mass asymmetric ($A_F\approx 208$) are of the 
same magnitude. The details of calculations are given below.
\section{The model}
\label{model}

We describe the fission process within the Langevin 
approach 
\cite{15} 
, i.e., by solving the equations for the time evolution of 
the shape of nuclear surface of fssioning system. For the shape 
parametrization we use that of two-center shell model (TCSM) 
\cite{16} 
with 4 deformation parameters $q_{\mu} =z_0/R_0, \delta_1, \delta_2, \alpha$. 
Here $z_0/R_0$ refers to the distance 
between the centers of left and right oscillator potentials with $R_0=1.2 
A^{1/3}$, $R_{0}$ being the radius of spherical nucleus with the mass 
number $A$. The parameters $\delta _i$, where $i=1,2$ describe the 
deformation of the right and left fragment tips. The fourth parameter 
$\alpha $ is the mass asymmetry and the fifth parameter of TCSM shape 
parametrization $\epsilon $ was kept constant, $\epsilon =0.35$, in 
all our calculations.

The first order differential equations (Langevin equations) for the time 
dependence of collective variables $q_{\mu }$ and the conjugated momenta 
$p_{\mu }$ are:

\belar{lange}
\frac{dq_\mu}{dt}&=&\left(m^{-1} \right)_{\mu \nu} p_\nu , \\
\frac{dp_\mu}{dt}&=&-\frac{\partial F(q,T)}{\partial q_\mu} - \frac{1}{2}\frac{\partial m^{-1} _{\nu \sigma} }{\partial q_\mu} p_\nu p_\sigma
 -\gamma_{\mu \nu} m^{-1}_{\nu \sigma} p_\sigma \nonumber\\
 &+&g_{\mu \nu} R_\nu (t), \nonumber
\end{eqnarray}

where the sums over the repeated indices are assumed. In Eqs.\req{lange} the $F(q,T)$ is the 
temperature dependent free energy of the system, and $\gamma _{\mu \nu }$ 
and (m$^{-1})_{\mu \nu }$ are the friction and inverse of mass tensors and 
g$_{\mu \nu }$ is the random force.

The free energy $F(q, T)$ is calculated as the sum of liquid drop deformation energy 
and the temperature dependent shell correction $\delta F(q, T)$. 
The damping of shell correction $\delta F(q, T)$ with the excitation energy is described in detail in \cite{17}. 
The single particle energies are calculated with the deformed 
Woods-Saxon potential 
\cite{18,19} fitted to the aforementioned TCSM shape 
parameterizations. It is to be noted the free energy is equal to potential 
energy at zero temperature.

The collective inertia tensor $m_{\mu \nu }$ is calculated within the 
Werner-Wheeler approximation \cite{20} 
and for the friction tensor $\gamma 
_{\mu \nu }$ we used the wall-and-window formula,  \cite{21,22}.

The random force $g_{\mu \nu }R_{\nu }(t)$ is the product of white noise 
$g_{\mu \nu }R_{\nu }(t)$ and the temperature dependent strength factors 
$g_{\mu \nu }$. The factors $g_{\mu \nu }$ are related to the temperature 
and friction tensor via the modified Einstein relation,
\[
g_{\mu \sigma } g_{\sigma \nu } =T^\ast \gamma _{\mu \nu } 
\,,\,\, {\rm with} \,\,\,T^\ast 
=\frac{\hbar \omega }{2}\,\,\coth \,\,\frac{\hbar \omega }{2T}\,\,,\,\,\,\,
\]
where $T^{\ast }$ is the effective temperature \cite{23}. The parameter $\omega 
$ is the local frequency of collective motion \cite{23}. The minimum of 
$T^{\ast }$ is given by $\hbar \omega /2$. 

The temperature $T$ in this context is related to the reaction energy $E_x$ and the internal excitation energy $E^{\ast }$ by,
\[
E^\ast =E_{gs} +E_x -\frac{1}{2}m^{-1}_{\mu \nu } p_\mu p_\nu 
-V_{pot} (q,T=0)=aT^2,
\]
where $V_{pot}$ is the potential energy and $a$ is the level density parameter. 
More details are given in our earlier publications, see \cite{24,25,26,14}.
Initially, the momenta $p_{\mu }$ are set to be equal to zero, and calculations are 
started from the ground state deformation. Such calculations are continued 
until the trajectories reach the "scission point", which was defined as the 
point in deformation space where the neck radius reaches the value $r_{neck}= 1$ fm.
\section{Numerical results}
\label{results}

In Fig. 1 we show the fission fragment mass distributions of super-heavy 
nuclei from $^{274}$Hs to $^{306}$122 as  function of fragment mass number 
$A_F$. As one can see, at $E_x=30$ MeV the shell structure is washed out and all considered here nuclei fission symmetrically. At excitation energy $E_x=10$ MeV the lighter superheavies Hs and Ds also undergo mass symmetric fission. The FFMDs of nuclei from $^{286}$Cn to $^{306}$122 have three or four peak structure. Obviously, the multi-peak structure of FFMDs is the result of shell effects, which at $E_x=10$ MeV are still large. 
\begin{figure}[ht]
\centering
\includegraphics[width=0.99\columnwidth]{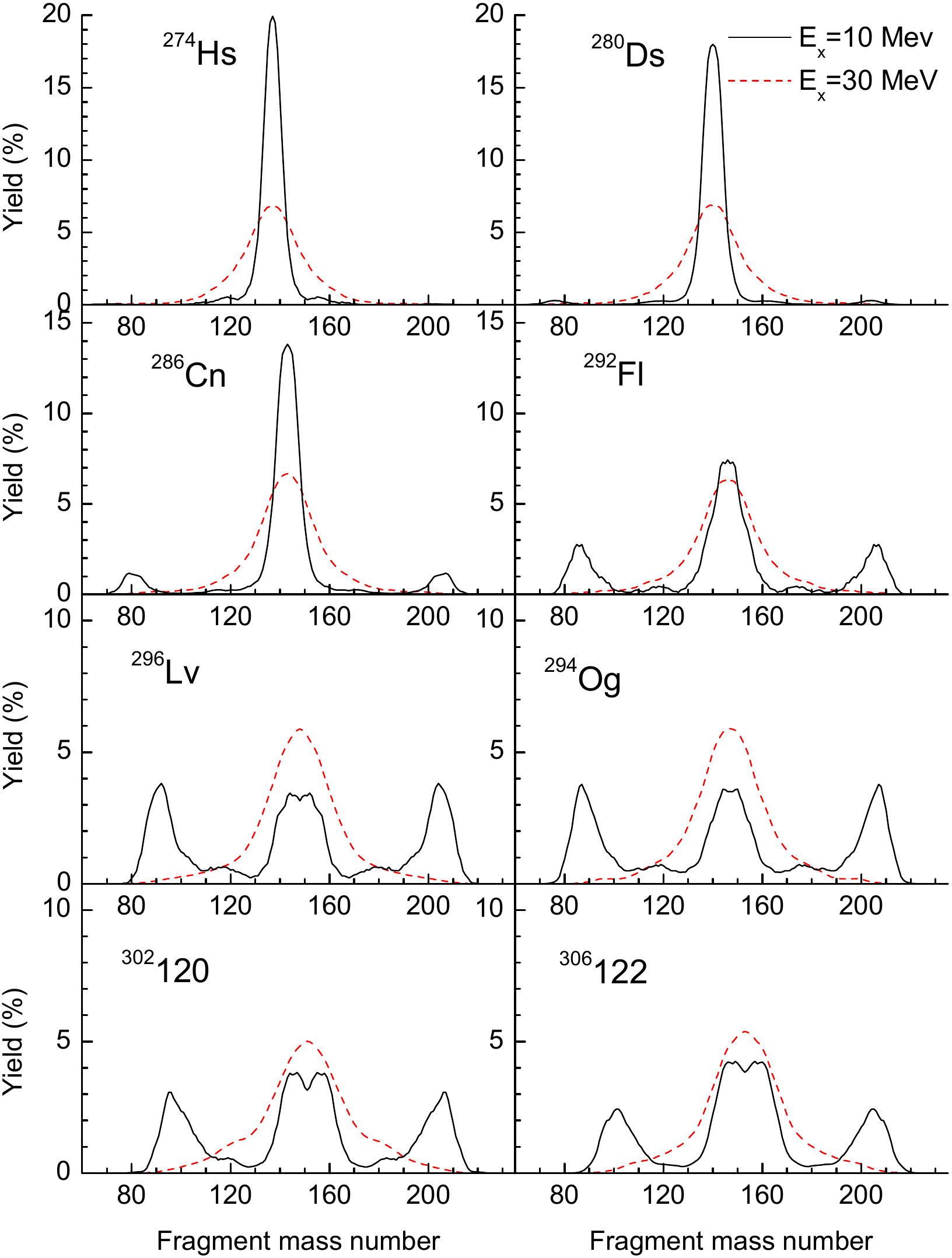}
\caption{The fission fragment mass distributions  of super-heavy nuclei from $^{274}$Hs to $^{306}$122 calculated for the excitation energy $E_x=10$ MeV and $E_x=30$ MeV as  function of fragment mass number}
\label{fig1}
\end{figure}
The symmetric peak which in heavier SHEs is split into two components. The peaks of lighter fragments are located around 
$A_F=140$. 

One can also see the strongly asymmetric peak at the mass number close to 
$A_F=208$. The strength of the (almost) symmetric and asymmetric 
components in FFMD of SHEs depends on the proton and neutron numbers of the 
compound nucleus. For $^{286}$Cn the contribution of strongly asymmetric 
peak is very small. This contribution becomes larger for more heavy SHE. In 
the element $^{306}$122 the symmetric and mass asymmetric peaks are of the 
same magnitude.

In order to understand the reason of such complicated structure we have looked at the potential energy of fissioning nuclei.
\begin{figure}[ht]
\centering
\includegraphics[width=0.75\columnwidth]{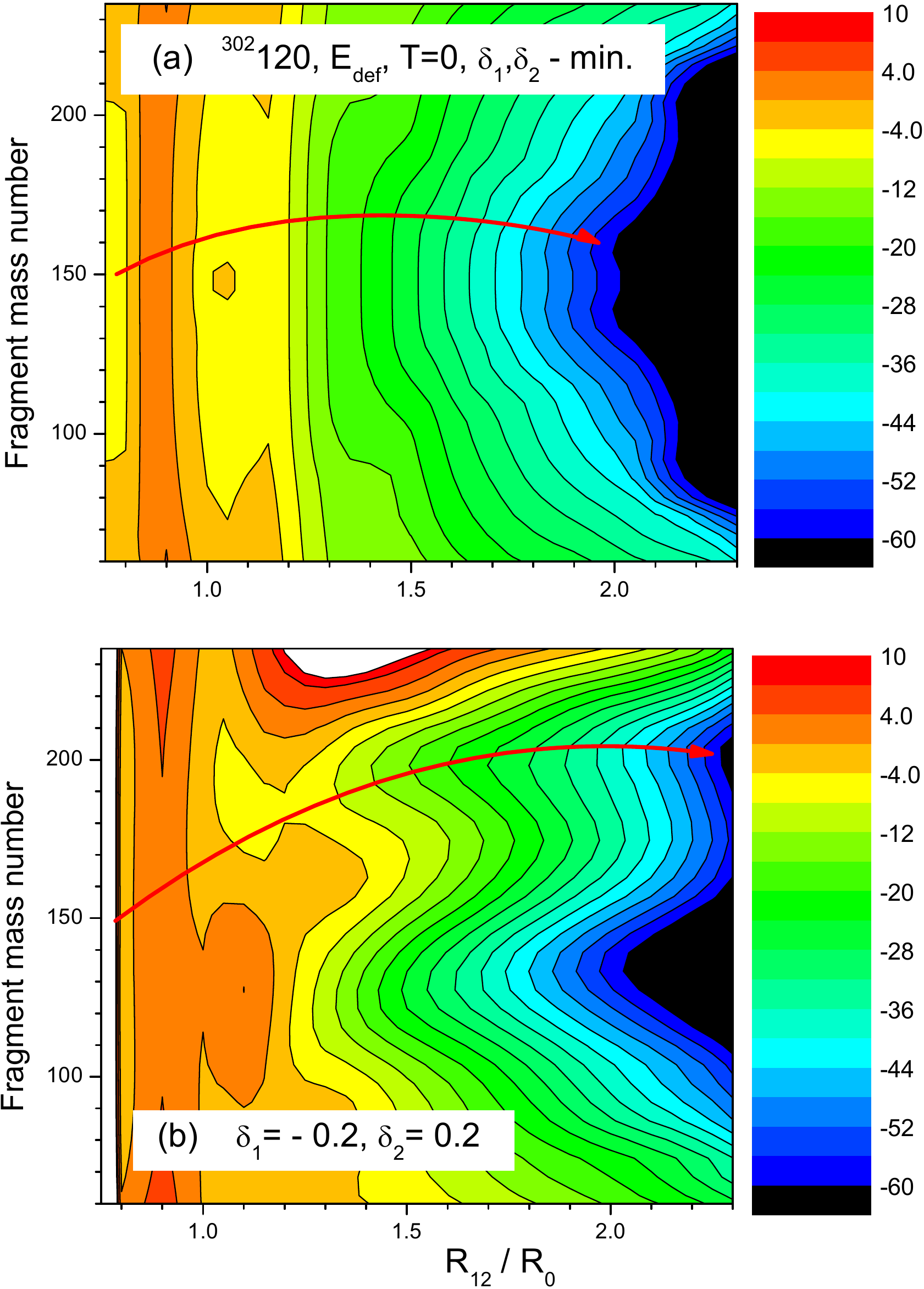}
\caption{(a) The potential energy of $^{302}120$ at $T=0$ minimized with respect to deformation parameters $\delta_1$ and $\delta_2$. (b) The potential energy of $^{302}120$ at $T=0$ at fixed values $\delta_1=-0.2$ and $\delta_2=0.2$.}
\label{fig2}
\end{figure}
Fig. 2 shows the potential energy $E_{def}$ of nucleus with $Z=120$ and $A=302$ 
at zero temperature as function of elongation (the distance $R_{12}$ between 
left and right parts of nucleus) and mass asymmetry. In  
Fig.2(a) the energy was minimized with respect to the deformation parameters 
$\delta _{1}$ and $\delta _{2}$. One clearly sees the bottom of 
potential energy leading to almost symmetric mass splitting. There is also a 
hint on the mass asymmetric valley at $A_F$ close to $A_F=208$. If the 
trajectories would follow the bottom of potential energy then the mass FFMD 
of $^{302}$120 would be mass symmetric. However it is well known that due to 
dynamical effects the trajectories may deviate substantially from the bottom 
of potential valley. We calculate the trajectories in four-dimensional 
deformation space. In this space there could be the local minima leading 
away from the bottom of potential valley. An example is shown in Fig. 2(b). 
Here we show the potential energy for fixed $\delta _1= 
-0.2$ and $\delta _2=0.2$.

One can see that in this subspace the trajectories can easily be trapped in 
the higher in energy valley leading to highly asymmetric fission. The 
trajectories can not skip into deeper symmetric valley because of barrier 
between these two valleys. In this way the strongly mass asymmetric peak 
appears in the mass distribution of fission fragments. In order to understand why this effects get stronger for heavier SHEs we 
have compared the dependence of potential energy close to the scission 
point, at $R_{12}/R_0=2.3$, on the mass asymmetry for two nuclei, 
$^{286}$Cn and $^{306}$122, see Fig. 3.

\begin{figure}[ht]
\centering
\includegraphics[width=0.95\columnwidth]{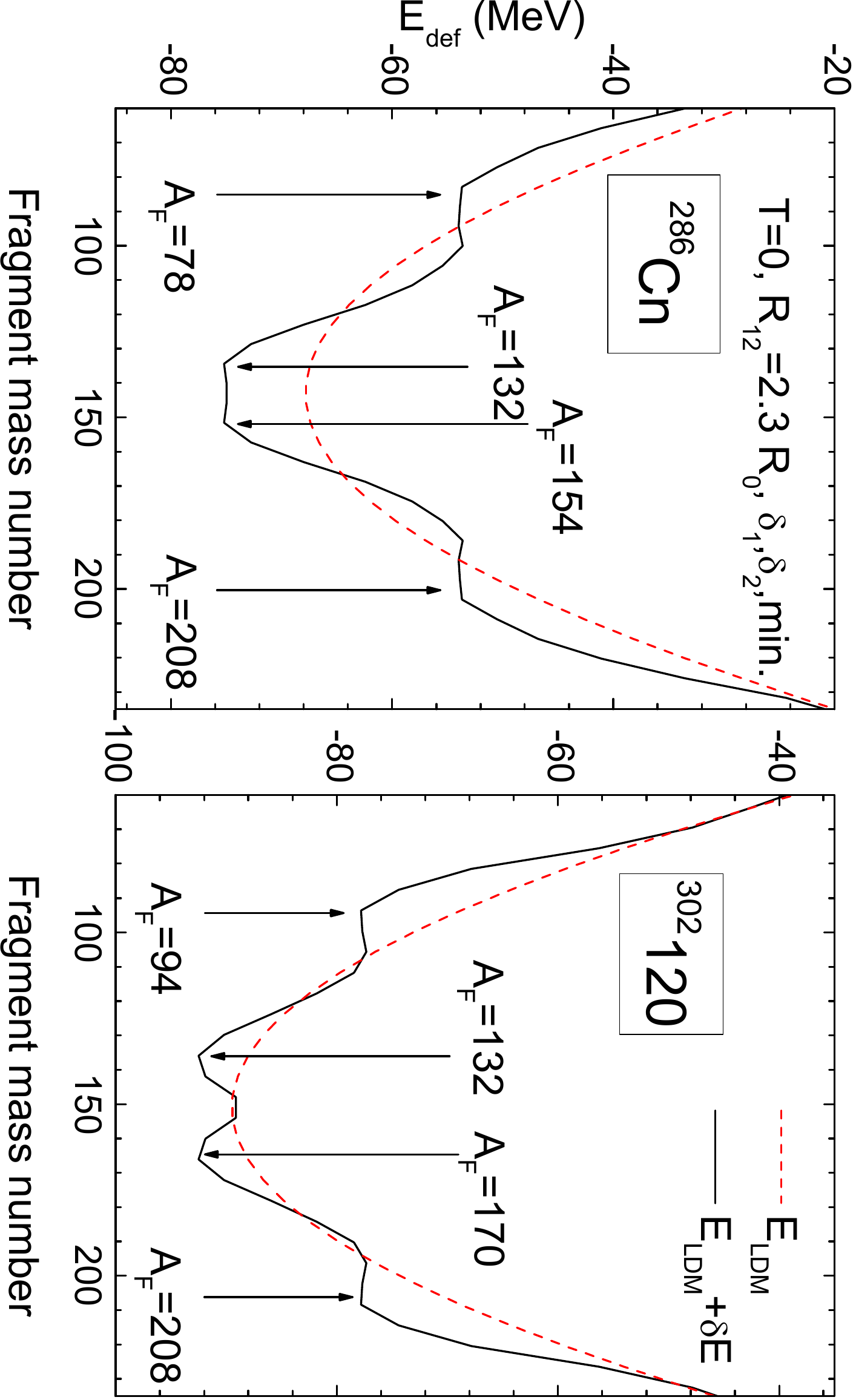}
\caption{The liquid drop (dash) and the total (solid) deformation energy near the scission line ($R_{12}=2.3 R_0$) for $^{286}$Cn and $^{302}$120 as function of fragment mass number.}
\label{fig3}
\end{figure}
In Fig.~\ref{fig3} we compare the total deformation energies near the scission line for $^{286}$Cn and $^{302}$120 with 
those of the liquid drop model. 
One can see that in case of $^{286}$Cn the local minima corresponding to 
$A_F=208$ and its paired fission fragment $A_F=78$ are by 20 MeV higher than the minimum around $A_F=132-154$. For 
$^{306}$122 the difference in almost symmetric and strongly mass asymmetric 
minima is smaller, only 14 MeV. Thus in this case the trajectories have more 
chances to get into the mass asymmetric valley at $A_F=208$ and its pair $A_F=94$.
As a result, the obtained FFMD becomes double mass asymmetric as seen in Fig.~\ref{fig1}.

Another reason for the appearing of $A_F=208$ contribution is the $Z/A$ 
ratio. The $Z/A$ ratio of fission fragment and of mother nucleus is 
approximately the same. For $^{132}$Sn this factor is equal to 0.379, while 
for $^{208}$Pb this ratio is equal to 0.394. The last ratio is much closer 
to that of $^{286}$Cn and $^{306}$122 which are equal to 0.392 and 0.397 
correspondingly. 

In Fig. 4 we investigate the quadrupole deformation $Q_{20}$ of the 
fragments, $Q_{20}=\langle r^{2 }Y_{20}(\cos \theta ) \rangle $ 
from $^{236}$U to $^{306}$122. 
The $Q_{20}$ is the main measure of the deformation of fragment's shape. The negative 
$Q_{20}$ corresponds to the oblate shape, the shape is spherical at 
$Q_{20}=0$, and positive $Q_{20}$ corresponds to the prolate shape. 
In actinides from $^{236}$U to $^{259}$Lr, there is no sign of $^{208}$Pb shell.
On the other hand, in SHEs from $^{274}$Hs to $^{306}$122 one can clearly observe 
the $Q_{20}(A)$ distributions located in both $A_F=132$ and 208,
though we hardly see the peak at $A_F=208$ of $^{274}$Hs in the mass distribution shown in Fig.~\ref{fig1}.
Note that the averaged $Q_{20}$ in actinides from $^{236}$U to $^{257}$Fm have positive $Q_{20}$ in common,
while the averaged $Q_{20}$ in actinides from $^{258}$Fm to $^{259}$Lr commonly have $Q_{20} \simeq 0$. 
It means that deformed shell around $A_F=132-140$ dominates the nuclear fission of actinides up to $^{257}$Fm,
though the spherical $^{132}$Sn strongly affects the fission of actinides at and above $^{258}$Fm.
In the same way, in SHEs, we found that the fragments, with the mass number $A_F=132-140$ 
are both deformed and of spherical shape with $Q_{20}\geq 0$,
the fragments with $A_F=208$, are spherical with $Q_{20}\approx 0$.
In this manner, we can demonstrate that two spherical magicities at $A=132$ and 208 play decisive roles in fission mechanisms.
\begin{figure}[ht]
\centering
\includegraphics[width=0.48\textwidth]{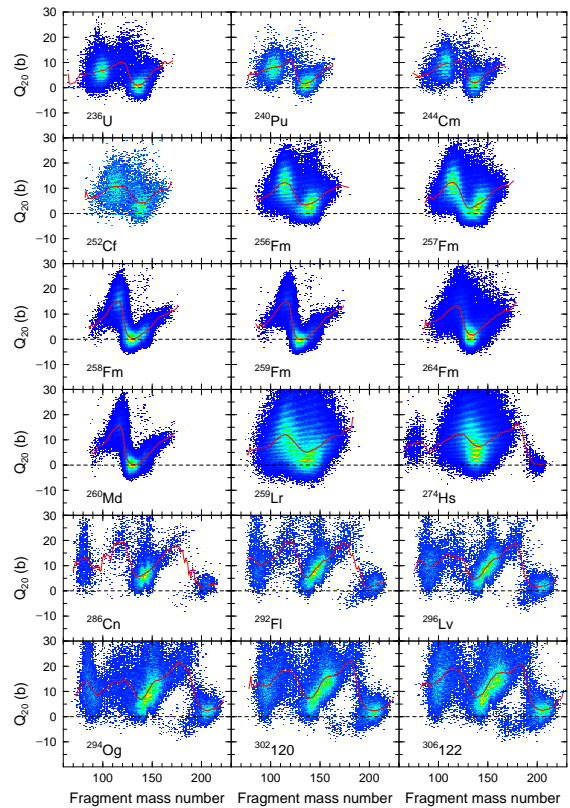}
\caption{The distribution of quadrupole deformation $Q_{20}$ with respect to fission fragment mass number for nuclei from $^{236}$U to $^{306}$122. The red curves mark the average values of $Q_{20}$.}
\label{fig4}
\end{figure}

Such results are quite reasonable because these fragments are nuclei with the double-closed shells. Another notable feature of $Q_{20}(A_F)$ plots 
is the difference of the distribution pattern between actinides ($^{236}$U 
to $^{259}$Lr) and SHN ($^{286}$Cn to $^{306}$122). The $Q_{20}(A_F)$ 
distributions of actinides consist of two groups; the nearly spherical heavy 
fragments with $A_F=132-140$ and the prolate light fragments. For the 
super-heavies from $^{286}$Cn to $^{306}$122, we observe the spherical heavier 
fragments with the mass number around $A=208$ and the complementary lighter 
fragments in addition to the mentioned above two groups seen in actinides. 

In \cite{14} we have noticed a very accurate correlation between the dependence 
of elongation of fragment and the multiplicity of prompt neutrons -- the 
number of neutrons per fission event emitted from completely accelerated 
fragments. 
So, the averages values of $Q_{20}$ (solid curves in Fig.4) represent actually the mass 
dependence of neutron multiplicity, what is an important observable of the 
fission process. 

It should be pointed out that, in the experiment by Itkis group \cite{6,7} they 
found a peak around $A=208$ and at complimentary light mass numbers. However, 
these peaks were assigned to be formed by quasi-fission process, not by 
fusion-fission.  
Such an interpretation is natural since the composite systems formed by hot-fusion reactions have excitation energy at least around 30 MeV, and the subtle shell effect, which gives rise to formation of the $A=208$ and complimentary fragments, is washed out.
Our calculation tells that only a small fraction of this peak can be  
indeed from fission of compound  nucleus (indicating SHE was formed with slightly larger 
probability), but it is overwhelmed by the quasi-fission component so it
cannot be identified in experiments. The  only possibility that this superasymmetric fusion-fission component can be observed is after emission of a few prescission neutrons to cool the residues to excitation energy region down to around 10 MeV.  If, e.g.,  multiplicities of prescission neutrons and fission fragments from corresponding residues are observed in coincidence, there is a chance that this superasymmetric component to be identified to come from fusion-fission events.  
It is highly desirable to have an 
experimental setup to distinguish these two components, namely, quasi-fission and fusion-fission, forming the same peaks.  

\section{Summary}
Within the 4-dimensional Langevin approach we have 
calculated the mass distributions of fission fragments of super-heavy nuclei 
from $^{274}$Hs to $^{306}$122. We have found a three-four peaks structure 
of mass distributions. In light super-heavies we see the dominant mass 
symmetric peak and small contributions from two highly asymmetric peaks at 
$A_H\approx 208$ and at the supplementary light fragment mass 
$A_L=A-208$. With increasing mass of fissioning nuclei the symmetric peak 
splits into two components and the strongly mass asymmetric peaks become 
higher. For $^{306}$122 all four peaks in FFMD are approximately of the same 
magnitude. So, the answer to the question: ``Fission of super-heavy 
elements: $^{132}$Sn-plus-the-rest, or $^{208}$Pb-plus-the-rest ? `` is: 
BOTH, the fragment with the mass number close to $^{132}$Sn, $A_F\approx 140$ 
plus the rest, and the fragment with the mass number $A_F\approx 208$ 
with spherical shape
plus the rest.  

{\bf Acknowledgments.}
 This study comprises the results of "Research and 
development of an innovative transmutation system of LLFP by fast reactors" 
entrusted to the Tokyo Institute of Technology by the Ministry of Education, 
Culture, Sports, Science and Technology of Japan (MEXT) and KAKENHI Grant 
Number 18K03642 from Japan Society for the Promotion of Science (JSPS). One of us (F. I.) was supported in part by the project  "Fundamental
research in high energy physics and nuclear physics" of the National Academy
of Sciences of Ukraine.
We appreciate very much the useful discussions with Prof. N. Carjan and Prof. A.V. Karpov.
 

\end{document}